\documentclass[sn-apa]{sn-jnl}% APA Reference Style

\jyear{2022}%

\usepackage{amssymb}
\usepackage{cleveref}
\usepackage{multirow}
\usepackage{graphicx}
\usepackage{epstopdf}
\usepackage{subfigure}
\usepackage{url}

\begin{document}

\title{Federated Graph Attention Network for Rumor Detection}

\author*[1]{\fnm{Huidong} \sur{Wang}}\email{huidong.wang@ia.ac.cn}

\author[1]{\fnm{Chuanzheng} \sur{Bai}}\email{whitecz@163.com}

\author[2]{\fnm{Jinli} \sur{Yao}}\email{jinli.yao@mail.concordia.ca}

\affil[1]{\orgdiv{School of Management Science and Engineering}, \orgname{Shandong University of Finance and Economics}, \orgaddress{\street{7366 East Erhuan Rd.}, \city{Jinan}, \postcode{250014}, \state{Shandong Province}, \country{China}}}

\affil[2]{\orgdiv{Concordia Institute for Information Systems Engineering}, \orgname{Concordia University}, \orgaddress{\street{1455 De Maisonneuve Blvd}, \city{Montreal}, \postcode{H3G 1M8}, \state{Quebec}, \country{Canada}}}

\abstract{With the development of network technology, many social media are flourishing. Due to imperfect Internet regulation, the spread of false rumors has become a common problem on those social platforms. Social platforms can generate rumor data in their operation process, but existing rumor detection models are all constructed for a single social platform, which ignores the value of cross-platform rumor. This paper combines the federated learning paradigm with the bidirectional graph attention network rumor detection model and proposes the federated graph attention network(FedGAT) model for rumor detection. Taking the advantages of horizontal federated learning for cross-platform rumor detection, the security of each social platform's data information can be ensured. We construct both the server-side and client-side models as a bidirectional graph attention network rumor detection model in the federated learning paradigm framework. The local model on the client-side can train and verify the rumor data of the social platform in the iterative process, while the model on the server-side only plays the role of aggregating the characteristics of different social platform models and does not participate in the training of the model. Finally, we conduct simulation experiments on the rumor datasets and the experimental results show that the federated graph attention network model proposed in this paper is effective for cross-platform rumor detection.}


\maketitle

\section{Introduction}\label{sec1}

The rapid development of network technology has made online social platforms an important place for us to obtain news and maintain social relationships in our daily lives. Many people reply to posts of interest when browsing the Internet, and then these replies also have corresponding comments. People can freely express their opinions on different media platforms and can forward or comment on the posts made by others. Nowadays, it is very common to share our lives or the latest news online, and the spread of the information is more convenient than before. When the content published on public media platforms cannot guarantee authenticity, it may cause the spread of rumors \cite{meel2020fake}. Rumors are news that is not validated. They can spread from person to person in a short period, and would seriously hinder people from obtaining true information. In addition, it may even cause huge economic losses or public panic in an emergency, thus the rumors should be detected as soon as they appear \cite{liang2015rumor}. Early rumor detection is conducive to the long-term stability of society.

The problem of rumor spreading on social networks has been researched by scholars in many fields such as sociology, journalism, communication, and management through different advanced methods. Many excellent detection models are proposed in recent years that can efficiently detect false rumor information. Some current rumor detection methods are achieved through empirical analysis, and some use the Decision Tree \cite{castillo2011information} or Support Vector Machine \cite{ma2017detect} in machine learning and other technologies for model processing. After the emergence of deep learning technology, with its advantages in processing large-scale data, various neural network models have been successfully applied to solve different problems of rumor detection, such as the difficulty of quick rumor detection, the sparseness of early rumor data, and the dynamic nature of rumor information. Among the rumor detection models based on neural networks, those using graph neural networks can fully consider the graph structure information constructed in the process of rumor propagation, so they can get great classification results \cite{providel2020using}.

At present, there are many social network platforms that have similar interactive functions for posting and replying to posts on the Internet, so similar user usage data can be generated in their daily activities, and there are similar rumor data features on those platforms \cite{wei2015exploiting}. Some false rumors may not appear on different social platforms simultaneously, if the platform can take the rumor features of other platforms to perform rumor detection on its own data, it can timely detect some new false rumors, and can effectively deal with quick rumor. For data security and privacy protection, data between different organizations are generally not interoperable, and they cannot be easily aggregated and processed, this inability to share data is called the isolated data island problem. To solve the problem of isolated data island, many scholars seek ways to train machine learning models without having to concentrate all the data in central storage, and federated learning was proposed. Federated learning is an emerging model framework proposed to solve the problem of isolated data island, and it can maximize the aggregation of data information from different institutions on the premise of ensuring data security and privacy. 

To solve the problem of isolated data island in rumor detection, we innovatively combine federated learning and graph neural networks to build a federated graph attention model for rumor detection. Our overall framework is to establish a rumor detection model based on the graph attention network and then effectively integrate the federated learning paradigm with the model. By adjusting the applicability of the model after introducing federated learning, cross-platform applications of the model can be realized, and the data features of different platforms can be fully extracted to achieve better rumor detection. Using the graph attention network can make full use of the graph structure information formed in the process of rumor dissemination and focus on the rumor nodes that have an important impact on the prediction results during the model training process. Our model can maximize the aggregation of the overall data information and enhance the usability of the model under the premise of ensuring the privacy and security of the data of different institutions, it can help multiple parties to build a shared high-performance rumor detection model. In addition, another advantage of applying federated learning is that it can reduce the computational pressure of the participants in rumor detection. By exchanging the intermediate calculation results of the model training process between multiple clients and servers, the computing power of the terminal device can be greatly utilized, the equipment requirements of each client organization can be reduced, and the model has better promotion value.

Our main contributions can be summarized in the following aspects. First, we study the problem of isolated data island when processing data from different platforms, and consider the way of cross-platform rumor data aggregation in the process of rumor detection. Second, to solve the problem of data islands in the rumor detection process, we introduce the horizontal federated learning framework into the rumor detection model based on the graph attention network, so that the rumor data of different platforms can be processed collaboratively while ensuring data security. Finally, we manually construct the public rumor dataset as different clients rumor datasets to simulate the FedGAT model in the experimental phase. It can be seen from the performance evaluation of the model experiment that the proposed cross-platform rumor detection model is effective, and it has significant advantages in addressing the isolated data island problem.

The main contents of the rest sections are arranged as follows. In Section II, we present introductions to the model for the rumor problem we address and the existing related research work on the federated learning paradigm. In Section III, we provide an overview of the tasks to be tackled and the theoretical knowledge used in our model. In Section IV, we describe the framework of our proposed FedGAT model in detail. The experiments and results are described in Section V. Finally, Section VI summarizes the research work of this paper.

\section{Related work}\label{sec2}

The rumor detection task has been studied by experts in many different fields, and there are many detection models based on different theoretical methods including game theory, empirical research, infectious disease models, machine learning, and so on. Since the research in this paper is based on the graph neural network, belonging to the field of deep learning, so we roughly divide the existing related research work based on whether they use deep learning technology or not. A brief overview of these research works are introduced below.

The spread of rumors has many similarities with the spread of the epidemic. Infectious disease models have corresponding applications in the early research on rumor detection, Devavrat Shah \cite{shah2011rumors} conducted a systematic study of the source of rumors, and he used infectious disease models to model the spread of rumors. He also constructed a rumor source estimator and probabilistically analyzed its threshold. Shang et al. \cite{shang2015epidemic} used the infectious disease model to consider how rumors spread among repeated populations in different social groups. To solve the problem of dynamic changes caused by the influence of rumors in the process of rumor-information confrontation, Xiao et al. \cite{xiao2019rumor} constructed a propagation dynamics model based on an evolutionary game to study the characteristics of rumor propagation. Nour El-Mawass \cite{el2020similcatch} used the supervised classifier to obtain the prior information of the probability graph framework, and then constructs the Markov random field and obtains the user's posterior prediction, which effectively alleviates the dynamic influence of the evolution of spam in rumor detection. Ma et al. focused on feature extraction of rumor texts, used an entity recognition method to enhance the understanding of rumor semantics, and applied ordinary differential equation network to detect rumors \cite{ma2021novel}. Some traditional machine learning methods for rumor detection include Support Vector Machine \cite{ma2017detect}, Naive Bayes model \cite{qazvinian2011rumor} usually require pre-defined text features, these models have strong data dependencey, and poor generalization ability. The research of machine learning in rumor detection now focuses on the deep neural network model.

Many classic deep learning neural network models have been applied in rumor detection. For example, Ma et al. proposed two different models for rumor detection based on Recurrent Neural Networks in 2016 \cite{ma2016detecting} and 2018 \cite{ma2018rumor}. Convolutional Neural Networks are widely used in the field of computer vision, Liu et al. \cite{zheng2017rumor} tried to use Convolutional Neural Networks for rumor detection, and constructed the rumor detection framework by combining word vectors with multi-layer convolutions. Wu et al. \cite{wu2020rumor} considered how to better perform feature transfer between interconnected post nodes in rumor information, and proposed a representation learning algorithm for rumor detection based on Gated Graph Neural Network. Based on Generative Adversarial Networks for rumor detection, Cheng \cite{cheng2021rumor} proposed a GAN-based hierarchical model for rumor detection with explanations, where the generator intelligently inserts controversial information into non-rumors to generate rumors and forces the discriminator to detect detailed Troubleshoot and accurately infer the problematic part of the sentence. Wang et al. constructed a sentiment lexicon to capture different human emotional responses, and then proposed a gated recurrent unit model based on sentiment lexicon and dynamic time series algorithm for rumor detection \cite{wang2020rumor}. Bian et al. \cite{bian2020rumor} applied Graph Neural Network to rumor detection, proposed a bidirectional graph convolutional neural network model with enhanced root node information, which considered the importance of the content information of the source post in the spread of rumors. Later, Bai proposed to convert the Graph Convolutional Network in this model into a Graph Attention Network \cite{bai2021rumor}, which can ignore the additional role of root node information and achieve better discrimination results.

Data has the attribute of assets, and more and more departments are paying attention to the value of data. Due to mutual competition, it is difficult for all parties to share data, which also leads to a state of data fragmentation. Based on this background, people began to seek a machine learning model that can be trained uniformly without having to concentrate all the data in a central storage point, thus resulting in federated learning. The core idea of our proposed model is to introduce the federated learning paradigm into the rumor detection model, and the research work on federated learning mainly focuses on improving security and dealing with statistical problems. Cheng et al. \cite{cheng2021secureboost} proposed the SecureBoost model for vertical federated learning, which is a boosted tree model dedicated to enhancing privacy protection. Liu et al. \cite{liu2020secure} proposed a federated transfer learning framework that can be flexibly applied to various multi-party security machine learning tasks. This framework enables knowledge to be transferred through transfer learning in the network without compromising user privacy. In the federated learning system, the participants that the model can assume can be divided into honest, semi-honest, or malicious. Bhagoji in his research \cite{bhagoji2019analyzing} shows the fragility of federated learning settings and advocates that there should be effective defense strategies to improve federated learning. At present, federated learning has been applied in many fields, such as medical image analysis in the field of computer vision \cite{sheller2018multi}. However, it has not been applied in the field of rumor detection, we innovate the related application model of rumor detection.

\section{Preliminaries}\label{sec3}

In this section, we first describe the task to be handled and then introduce the structure of the Graph Attention Network applied in this paper. Finally, we will introduce the federated learning paradigm and analyze how to combine this paradigm with Graph Neural Networks for rumor detection. 

\subsection{Problem statement and notation}\label{sec31}

It is difficult for multiple social platforms to work together to deal with the problem of rumor spreading. The traditional rumor detection research was conducted on a single online social network platform, usually just obtaining user data information of a certain organization and then building a model to determine whether a event post false rumor post or not. The rumor detection model constructed in this paper focuses on the interoperability of rumor data between different social platforms and establishes a model that can aggregate cross-platform data to improve the efficiency of detecting false rumor events in social networks. The definitions of some notations used in this paper are summarized in Table \ref{Table1} and the rumor detection problem to be dealt with is described as follows.

\begin{table}[h]
\centering
\begin{minipage}{210pt}
\caption{Notation summarization}
\label{Table1}
\renewcommand{\arraystretch}{1.5}
\setlength{\tabcolsep}{0.7mm}{
\begin{tabular}{c|c} 
 \hline
 \textbf{Notation} & \textbf{Definition} \\ 
 \hline
 $D$ & rumor data of a social platform \\ 
 \hline
 $C$ & the content of a rumor event \\ 
 \hline
 $Graph$ & rumor graph-structured data \\ 
 \hline
 $V$ & set of vertexes in $Graph$ \\ 
 \hline
 $E$ & set of edges in $Graph$ \\ 
 \hline
 $h$ & node features in $Graph$ \\ 
 \hline
 $m$ & total number of participating social platform clients \\ 
 \hline
 $k$ & the number of clients participating in the training \\ 
 \hline
 $F$ & local model on the client-side \\ 
 \hline
 $G$ & global model on the server-side \\ 
 \hline
\end{tabular}}
\end{minipage}
\end{table}

There are two or more online social platforms, and they all have their user usage data. We hope that when conducting rumor detection, each platform can not only use its unique rumor data for model analysis but also can obtain parameter information of model training from the other platforms, which can dig out more valuable features for rumor detection, which is helpful to improve the accuracy of model detection. Assuming there are $k$ platform participants, and their rumor data set is $\{ {D_i}\} _{i = 1,...,k}^m$, where ${D_i}$ denotes the data from the $i$-th platform, and the data is composed of post events published on this platform one by one. If the $i$-th platform data contains $m$ post events, it can be represented by a set $\{ {C_1},{C_2},...,{C_m}\}$. Each $C$ contains the content of the source post and the content of all other users' replies to this post, as well as the connection relationship between the replies and the post. The goal of rumor detection is to classify whether the content of the source post of an event is reliable, that is, to judge whether it is false rumor based on existing knowledge.

\subsection{Graph Attention Network}\label{sec32}

The Graph attention network is a model of graph neural network which introduces an attention mechanism. Compared with other graph neural network models, its main advantage is that it can assign different weights to the information of adjacent nodes on the graph, that is, it can focus on the more valuable graph nodes. In the graph structure composed of rumor data, the importance of the source post information is higher than that of other sub-nodes. Therefore, Graph Attention Network is very suitable for graph-structured data generated from rumor data.

Assuming that the feature of the $i$-th node on the graph is represented by ${h_i}$, and we defined $a$ as a simple feedforward neural network, the correlation coefficient between node $i$ and node $j$ can be calculated by $a[W \cdot {h_i}\Vert W \cdot {h_j}]$. We can further normalize the calculation results as shown in \cref{eq1}, where ${\alpha _{ij}}$ is the attention coefficient between two nodes and it is the core parameter in Graph Attention Network, ${N_i}$ is the set of all neighboring nodes of $i$-node and $W$ is the weight matrix which can be optimized during model training.

\begin{equation}
{\alpha _{ij}} = \frac{{\exp (LeakyReLU({a^T}[W \cdot {h_i} \Vert W \cdot {h_j}]))}}{{\sum\nolimits_{j \in {N_i} \cup i} {\exp (LeakyReLU({a^T}[W \cdot {h_i} \Vert W \cdot {h_j}]))} }} \label{eq1}
\end{equation}

In the model proposed in this paper, we additionally use the multi-head attention mechanism. When obtaining new node features, we repeat the steps of calculating the attention coefficients $Heads$ times, and perform averaging or concentration operations based on these results from $Heads$ times calculation. The new node feature $h_i^{'}$ calculation formula is shown in \cref{eq2}, where $||$ denotes the concatenation operation and $\sigma$ is a nonlinear sigmoid function.

\begin{equation}
h_i^{'} = \mathop \parallel \limits_{head = 1}^{Heads} \sigma (\sum\nolimits_{j \in {_i} \cup i} {\alpha _{ij}^{head}{W^{head}}{h_j}} ) \label{eq2}
\end{equation}

\subsection{Federated Learning}\label{sec33}

Federated learning aims to establish a federated learning model based on distributed data sets. It generally includes two processes: model training and model inference. In model training, model-related information can be exchanged between parties. Federated learning is an algorithm framework used to build machine learning models with the following characteristics. First, two or more federated learning participants collaborate to build a shared machine learning model, and each participant has several training data that can be used to train the model. Second, in the training process of the federated learning model, the data owned by each participant will not leave the participant, that is, the data will not leave the data owner. Information related to the federated learning model can be transmitted and exchanged between parties in an encrypted manner, and it is necessary to ensure that not a participant can infer the original data of other parties. In addition, the performance of the federated learning model must be able to fully approximate the performance of the ideal model which means the machine learning model obtained by gathering and training all training data.

We use horizontal federated learning, which processes cross-platform data with the same feature $X$ and label information $Y$ while the sample data is different, so it works for the rumor detection situation we studied. In the typical federated learning paradigm, the local objective function of the i-th client is shown in \cref{eq3}, where ${D_i}$ is the local dataset of the $i$-th client, $f$ is the loss function of the model with parameter $w$, and $n_i$ is the data volume of the i-th client \cite{yang2019federated}.

\begin{equation}
{F_i}(w) = \frac{1}{{{n_i}}}\sum\limits_{j \in {D_i}} {f{}_j(w)} \label{eq3}
\end{equation}

The central server objective function $F(w)$ is usually calculated as \cref{eq4}, where $m$ is the total number of client devices participating in the training and $n$ is the sum of all client data volumes.

\begin{equation}
\mathop {\min }\limits_w F(w) = \sum\limits_{i = 1}^m {\frac{{{n_i}}}{n}{F_i}(w)} \label{eq4}
\end{equation}

\section{FedGAT model}\label{sec4}

\begin{figure}[h] 
\centering  
\includegraphics[width=120mm]{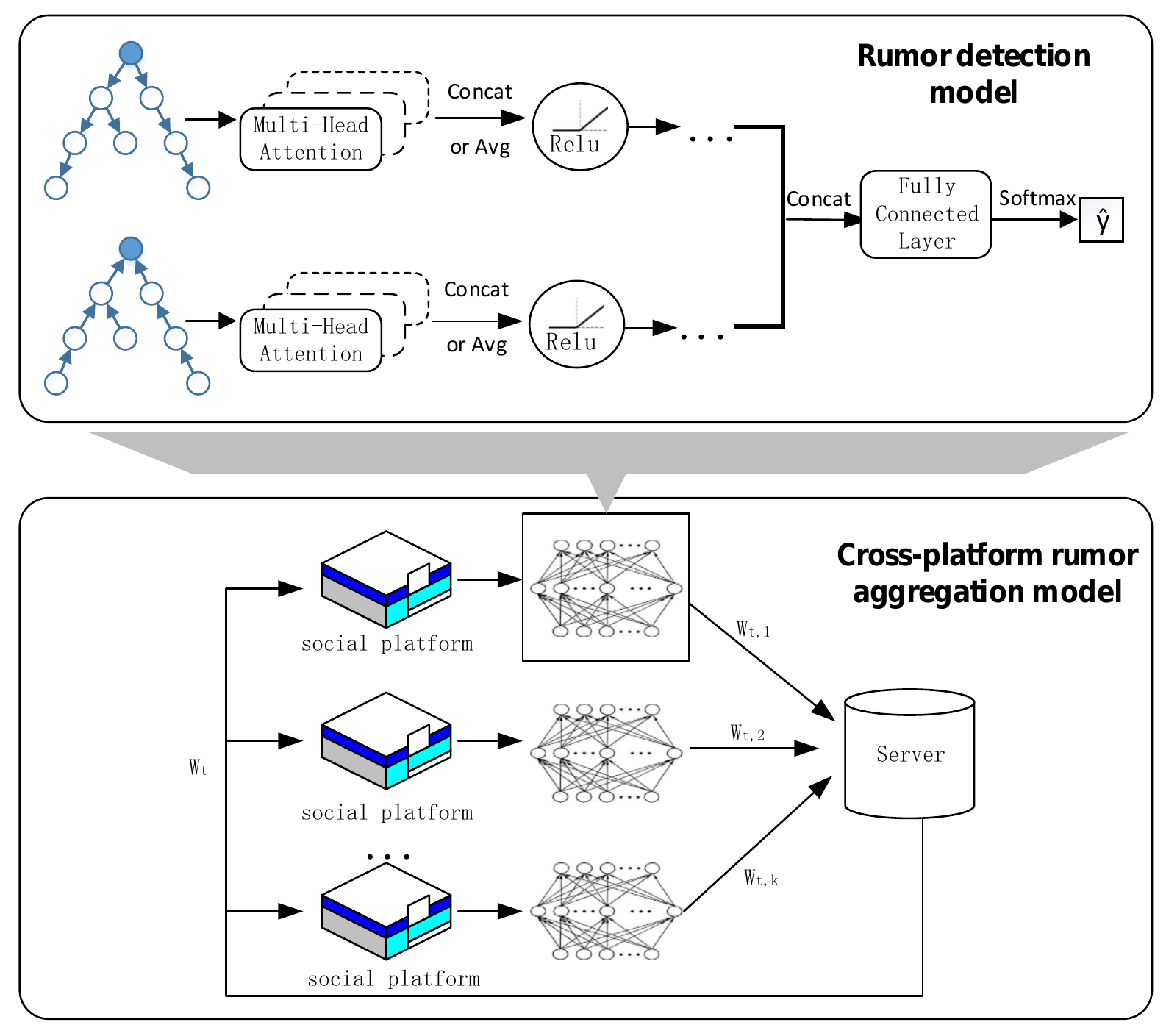}
\caption{Fed-GAT rumor detection model.}  
\label{fig:model}
\end{figure}

In this paper, we extend the Bidirectional Graph Attention Network model applied to handle rumor detection on a single social platform into the framework of the federated learning paradigm, making it applicable to cross-platform rumor detection. The overall framework of the proposed FedGAT (federated graph attention network) model is shown in Fig.\ref{fig:model}. Each social platform uses proprietary data locally for rumor detection, and the model parameters generated during the detection model training process will be sent to the terminal server for aggregation processing. The global model on the server side will return the processed model parameters to the local models, and then the local model will fine-tune the parameters of its own model according to the cross-platform information. The cross-platform rumor detection model proposed in this paper consists of two parts, which are the local rumor detection model and the federated learning paradigm for cross-platform data processing, we will introduce our FedGAT model in detail below.

\subsection{Preprocessing of rumor data}\label{sec41}

From our daily experience, we know that the content information of a post can be reflected by the content of the parent and child posts connected to it. In addition, the original content information of the first post in the event is often the most important, and the subsequent posts are comments on the content of the original post. Therefore, we choose the Bidirectional Graph Attention Network as the basis for our rumor detection model, in which the two-directional model can synthesize the spread of rumor information along the top-down direction and the bottom-up direction, and the GAT model can enhance the attention to the source post information. Therefore, we first need to process the rumor data as bidirectional graph structure data and perform feature processing on the rumor text information.

Firstly, the text information in the rumor data is processed by word vector, and it is processed into a numerical type that is convenient for training according to the frequency of words in the text content in terms of TF-IDF. The dimension of the constructed word vector is 5000, and the value of each dimension represents the frequency of a certain word in the post. Then, the graph structure data $Grap{h_i} = ({V_i},{E_i})$ is constructed according to the retweet or response relationships in $i$-th rumor event, where the nodes $v \in {V_i}$ are the posts posted by the users on the social platform, and the edges $e \in {E_i}$ on the graph are the connection relationships between the posts. In the bidirectional graph structure data $Graph_i^{TD} = ({V_i},E_i^{TD})$ and $Graph_i^{BU} = ({V_i},E_i^{BU})$, the adjacency matrix ${A_i}$ is used to represent the connection relationship of nodes on the graph, and $A_i^{TD} = {(A_i^{BU})^T}$. The feature matrix ${H_i} = [h_{i,0}^T,h_{i,1}^T,...,h_{i,{n_i}}^T]$ contains the word vector processed for each post, and the feature matrix of the graph-structured data in both directions is the same. 

\subsection{Local model training on social platform}\label{sec42}

Social platforms participating in the horizontal federated learning each have a local rumor detection model and they can use their proprietary rumor data for local training. For a rumor event, it is described below how the operation on it is performed. It can be seen from the Fig.\ref{fig:model} that the rumor detection model processing processes in the two directions are similar. During the processing of the GAT model, the graph structure does not change, that is, the adjacency matrix $A$ of the rumor graph data is a fixed value during the model training process, and the feature matrix $H = [h_0^T,h_1^T,...,h_{{n_i}}^T]$ is always changing in the model, so the $m$-th node feature $h_m$ is constantly aggregating the information of other nodes.After obtaining the graph structure data, a two-layer multi-head GAT model is used in both directions and we set the amount of attention heads in the graph attention layer as 5, the multi-head attention results are concatenated at the first layer, and the average operation is used at the second layer. Nonlinear processing is performed using the \textit{ReLU} activation function in both layers, and the operations of the $i$-th post feature are shown in \cref{eq5} and \cref{eq6},
\begin{equation}
h_i^{TD{\rm{ }}'} = ReLU(\mathop {{\rm{||}}}\limits_{head = 1}^5 \sigma (\sum\limits_{j \in {_i} \cup i} {\alpha _{ij}^{TD{\rm{ }}head}{W^{TD{\rm{ }}head}}h_{_j}^{TD}} )) \label{eq5}
\end{equation}
   
\begin{equation}
h_i^{{\rm{BU }}'} = ReLU(\mathop {{\rm{||}}}\limits_{head = 1}^5 \sigma (\sum\limits_{j \in {_i} \cup i} {\alpha _{ij}^{{\rm{BU }}head}{W^{{\rm{BU }}head}}h_{_j}^{BU}} )) \label{eq6}
\end{equation}
where $||$ denotes the concatenation operation at the first layer and average operation at the second layer. Then, we take full connection processing on the concatenated results from top-down and bottom-up directions, and finally we use the \textit{softmax} function to process the output classification results as shown in \cref{eq7}.
 
\begin{equation}
\hat y = softmax(FC(||({H^{TD'}},{H^{BU'}})))
\label{eq7}
\end{equation}

\subsection{Aggregate model parameters on terminal server}\label{sec43}

The terminal server of horizontal federated learning is to aggregate the model training data of different social platforms, and after updating the global model, the updated parameter changes are sent back to those local social platforms. In the configuration here, we set the configuration files such as the number of clients for each round of training and the number of iterations, and take the previously defined Bi-GAT rumor detection model as the initial model on the server-side. This model is used to receive all client training parameter information.

The aggregation function we choose in this part is the classic FedAvg algorithm, and its calculation formula is shown in \cref{eq8},
\begin{equation}
{G^{t + 1}} = {G^t} + \frac{{\rm{1}}}{m}\sum\limits_{i = 1}^m {(F_i^{t + 1} - {G^t})}
\label{eq8}
\end{equation}
where $G$ and $F$ represent the global model and the local model respectively, $t$ refers to the $t$-th round of training. Its main function is to use the received model uploaded by the clients to update the global model after the constructor is defined.

\subsection{Local model parameter adjustment on social platform}\label{sec44}

After parameter update in the server, the social platforms will adjust its local model according to the returned parameter information. We copy the configuration information set on the server-side to the local. After receiving the optimized parameters of the global model from the server, the rumor detection model of each social platform client will be modified as shown in \cref{eq9},
\begin{equation}
F_i^{t + 1} = (1 - \lambda )F_i^t + \lambda {G^t}
\label{eq9}
\end{equation}
where $\lambda$ is a hyperparameter used to represent the degree to which other social platform data influences its local model, and the more similar the multiple social platforms participating in federated learning are, the larger the value of $\lambda$ is.

\section{Experiments}\label{sec5}

Due to the inconvenience of obtaining rumor datasets of different social platforms in simulation experiments, we chose two public rumor detection datasets of Twitter 15 and Twitter 16 \cite{ma2017detect} for model detection. The basic information of the two datasets is shown in Table \ref{Table2}. It can be seen that there are four types of sample labels for rumor data which are true rumor, false rumor, non-rumor, and unverified rumor (TR, FR, NR, UR). The distribution in these two datasets is uniform and there is no imbalanced data. When using the model proposed in this paper to detect rumors, only the detected rumors with the FR label indicate that there is a problem with this event, and the management department of the social platform needs to deal with the rumors promptly.

\begin{table}[h]
\centering
\begin{minipage}{210pt}
\caption{Statistics of Twitter 15 and Twitter 16 datasets}
\label{Table2}
\renewcommand{\arraystretch}{1.5}
\setlength{\tabcolsep}{1.5mm}{
\begin{tabular}{c|c|c} 
 \hline
 \textbf{Statistic} & \textbf{Twitter15} & \textbf{Twitter16} \\ 
 \hline
 Number of non-rumors & 374 & 205 \\ 
 Number of false rumors & 370 & 205 \\
 Number of true rumors & 372 & 207 \\
 Number of unverified rumors & 374 & 201 \\
 Total number of events & 1490 & 818 \\ 
 \hline
\end{tabular}}
\end{minipage}
\end{table}

The model implementation in this paper is based on the \textit{PyTorch Geometric Library} \cite{fey2019fast}, an extension library based on \textit{PyTorch} that facilitates writing and training graph neural networks. Due to the random nature of neural network training, we repeated the experiment and averaged the best results each time to evaluate the performance of the model.

\subsection{Comparative analysis of rumor detection experiments}\label{sec51}

When conducting experiments, the first thing we do is artificially construct rumor data from different social platforms. Since Twitter 15 and Twitter 16 are rumor data from Twitter processed similarly, we assume that these two datasets are from two different social platforms and we take them as input to our federated graph attention network for simulation experiments. After the federated learning paradigm is introduced, the social platform will update the features of the local model according to the rumor data from other platforms. To verify the effectiveness of the federated graph attention network proposed in this paper, we select some baseline models for comparison. The following are our our experimental rumor detection models:

\begin{enumerate}[1.]
  \item TD-RvNN \cite{ma2018rumor}: It is a rumor detection model based on the recurrent neural network using the GRU gate unit, and this model introduced the top-down tree.
  \item BU-RvNN \cite{ma2018rumor}: It is a variant of TD-RvNN, which differs from the previous model in that it introduced the bottom-up tree.
  \item Bi-GCN \cite{bian2020rumor}: The rumor detection model is based on the bidirectional graph convolutional neural network, and the importance of root node information is strengthened in the model.
  \item Bi-GAT \cite{bai2021rumor}: The rumor detection model is based on a bidirectional graph attention neural network, which assigns different weights to different nodes in the graph structure.
 \item FedGAT: Our proposed federated graph attention network model for cross-platform rumor detection task.
\end{enumerate}

In the simulation experiment, our model trains and verifies the rumor data of Twitter15 and Twitter16 on two clients respectively. The local model on the client-side is trained twice, and the value of $\lambda$ is set to 0.2. The server side aggregates the validation dataset of the two clients, and only validates the model without training the rumor data. It should be noted that, unlike the previous rumor detection models, our model has both local model and global model training during processing, that is, if the numbers of local model and global model training epochs are set to $a$ and $b$ respectively, then the actual number of training epoch is $a \times b$ times for each social platform. The training epoch times mentioned later in this paper are $b$ of the global model. Therefore, the FedGAT model will reach the optimal effect after training about 10 times, and the continuous training will overfit. We train the FedGAT model by minimizing the cross-entropy loss of the predicted and actual labels for all rumor events $C$, and use Adam Optimizer to update the parameters in our network model.In this experiment, the number of training times $b$ of the server-side global model is set to 15 times, which is far less than other traditional neural network models.

The main evaluation metric chosen in our experiments is the accuracy rate (Acc), which is calculated by dividing the number of correct predictions by the total number of events. In addition, the F1 scores of the four categories of the Twitter rumor dataset are used as an auxiliary reference. F1 is a commonly used evaluation indicator when evaluating the validity of the classification, it is the harmonic average of $Recall$ and $Precision$, as shown in \cref{eq10}. The evaluation metrics we chose are that the higher the value, the better the effect of the model.

\begin{equation}
F1= \frac{2 \cdot Precision \cdot Recall}{Precision + Recall}
\label{eq10}
\end{equation}

The experimental results using different detection models for Twitter 15 and Twitter 16 datasets are shown in Table \ref{Table3} and Table \ref{Table4}, and all optimal values are marked in bold. Among them, our FedGAT model is trained on these two datasets at the same time, and other comparative rumor detection models can only process one dataset at a time, so simulation experiments are carried out on these two datasets respectively.

\begin{table}[h]
\centering
\begin{minipage}{210pt}
\caption{Results for dataset Twitter15}
\label{Table3}
\renewcommand{\arraystretch}{1.5}
\setlength{\tabcolsep}{1.5mm}{
\begin{tabular}{l|l|l|l|l|l}
\hline
\multirow{2}*{\textbf{Method}} & \multirow{2}*{\textbf{Acc}} & \textbf{NR} & \textbf{FR} & \textbf{TR} & \textbf{UR} \\ \cline{3-6}
~ & ~ & \textbf{F1} & \textbf{F1} & \textbf{F1} & \textbf{F1}  \\ \hline
TD-RvNN & 0.7196 & 0.6667 & 0.7023 & 0.7920 & 0.7170  \\ \hline
BU-RvNN & 0.6686 & 0.6374 & 0.6666 & 0.7865 & 0.5940  \\ \hline
Bi-GCN & 0.8177 & 0.7605 & 0.8259 & \textbf{0.8948} & 0.7784  \\ \hline
Bi-GAT & 0.8315 & 0.8012 & 0.8268 & 0.8871 & 0.8009  \\ \hline
FedGAT & \textbf{0.8409} & \textbf{0.8092}  & \textbf{0.8500}  & 0.8152  & \textbf{0.8091}  \\ \hline
\end{tabular}}
\end{minipage}
\end{table}

\begin{table}[h]
\centering
\begin{minipage}{210pt}
\caption{Results for dataset Twitter16}
\label{Table4}
\renewcommand{\arraystretch}{1.5}
\setlength{\tabcolsep}{1.5mm}{
\begin{tabular}{l|l|l|l|l|l}
\hline
\multirow{2}*{\textbf{Method}} & \multirow{2}*{\textbf{Acc}} & \textbf{NR} & \textbf{FR} & \textbf{TR} & \textbf{UR} \\ \cline{3-6}
~ & ~ & \textbf{F1} & \textbf{F1} & \textbf{F1} & \textbf{F1} \\ \hline
TD-RvNN & 0.7423 & 0.6923 & 0.7160 & 0.8395 & 0.7209 \\ \hline
BU-RvNN & 0.7041 & 0.6000 & 0.7123 & 0.8889 & 0.6383 \\ \hline
Bi-GCN & 0.8497 & 0.7591 & 0.8478 & 0.9197 & 0.8768 \\ \hline
Bi-GAT & 0.8699 & \textbf{0.8096} & 0.8728 & 0.9318 & 0.8558 \\ \hline
FedGAT & \textbf{0.8906} & 0.79925 & \textbf{0.8763} & \textbf{0.9590} & \textbf{0.8845} \\ \hline
\end{tabular}}
\end{minipage}
\end{table}

According to the experimental results in Table \ref{Table3} and Table \ref{Table4}, we can observe that the FedGAT model proposed in this paper achieves the highest accuracy after the two datasets are input into the model at the same time, and the F1 values of most of the four categories are better than the corresponding results of other models. Therefore, the model proposed in this paper is effective in cross-platform rumor detection. In addition, we also find that the performances of the rumor detection models based on the graph attention network are better than that of other neural networks, and that is because the graph attention network model can pay more attention to the influential rumor data nodes when dealing with the graph structure data formed by rumor propagation, thereby improving the model detection performance.

\subsection{Parameter analysis}\label{sec52}

After the experimental comparative analysis part, we verify the effectiveness of the proposed cross-platform rumor detection model. We perform sensitivity analysis on some parameters in this model below. When a large number of social platforms are involved in rumor detection at the same time, some social platforms can not to participate in the FedGAT model training in each training epoch which is to solve the problem that not all platforms are always online in practice. That is, if $m$ social platforms participate in the rumor detection, only $k$ platforms are chosen to participate in the training in each training epoch. So, we first conduct an experimental analysis of the effect of parameters $m$ and $k$ on FedGAT model performance.

We manually split the Twitter15 and Twitter16 datasets into several independent subsets as client-side data for the FedGAT model to simulate real-life datasets from different social platforms. The local model parameter $\lambda$ in this part is set to 1, so the parameter values of each social platform in the next round of model training are equal to the aggregation of the parameter values of all participant models, which can be obtained by the \textit{FedAvg} algorithm on the global model. We set $m \in (1,2,4,10)$ and different $k$ values for comparative analysis when we take different $m$. In addition, when the number of clients $m$ and $k$ in the configuration file are both set to 1, it will be transformed into centralized learning, which is the simple Bi-GAT rumor detection model, and there is no cross-platform rumor information transmission.

\begin{figure}[htbp]
\subfigure
{
    \begin{minipage}[b]{.5\linewidth}
        \includegraphics[scale=0.42]{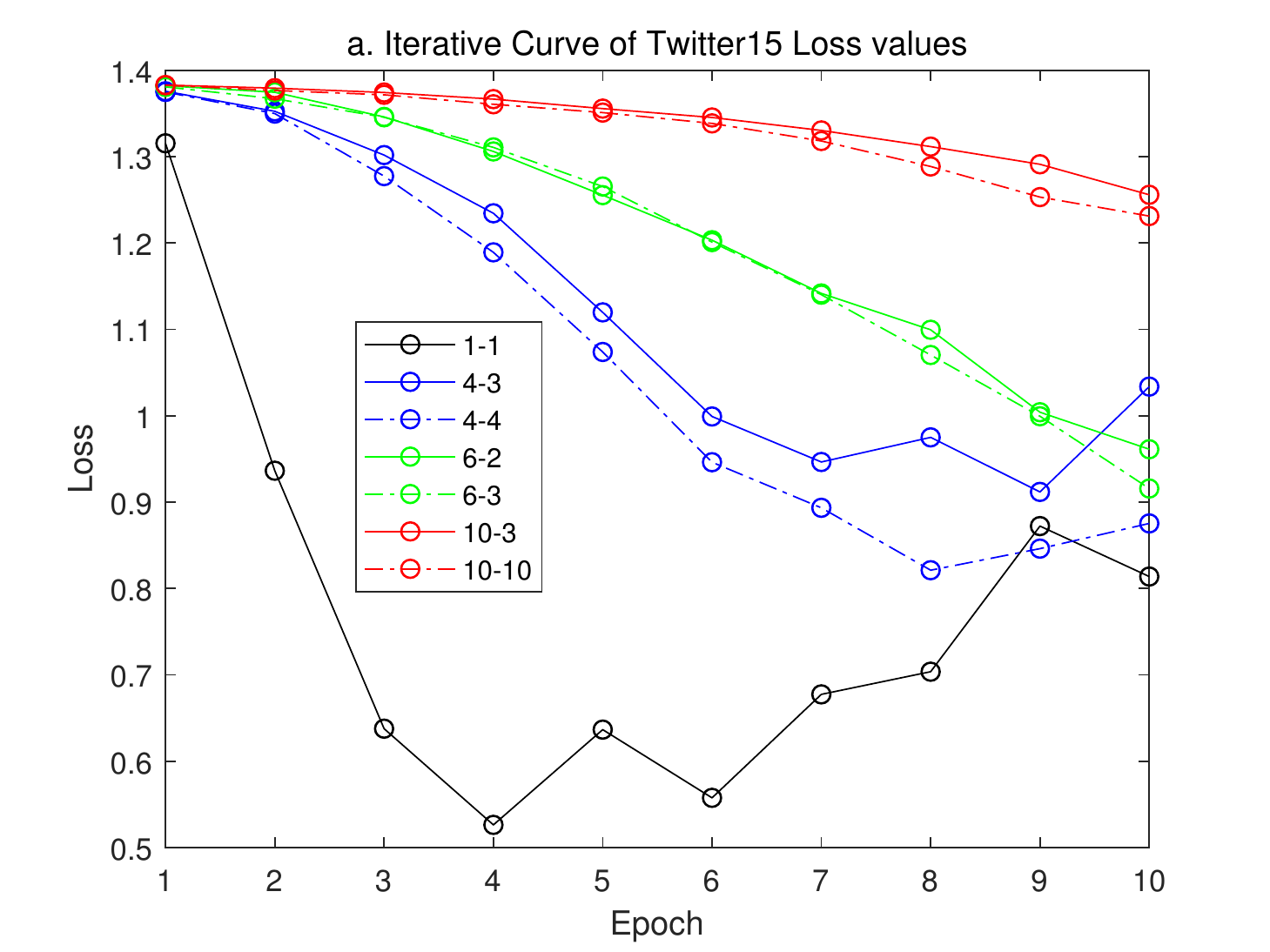}
    \end{minipage}
}
\subfigure
{
 	\begin{minipage}[b]{.8\linewidth}
        \includegraphics[scale=0.42]{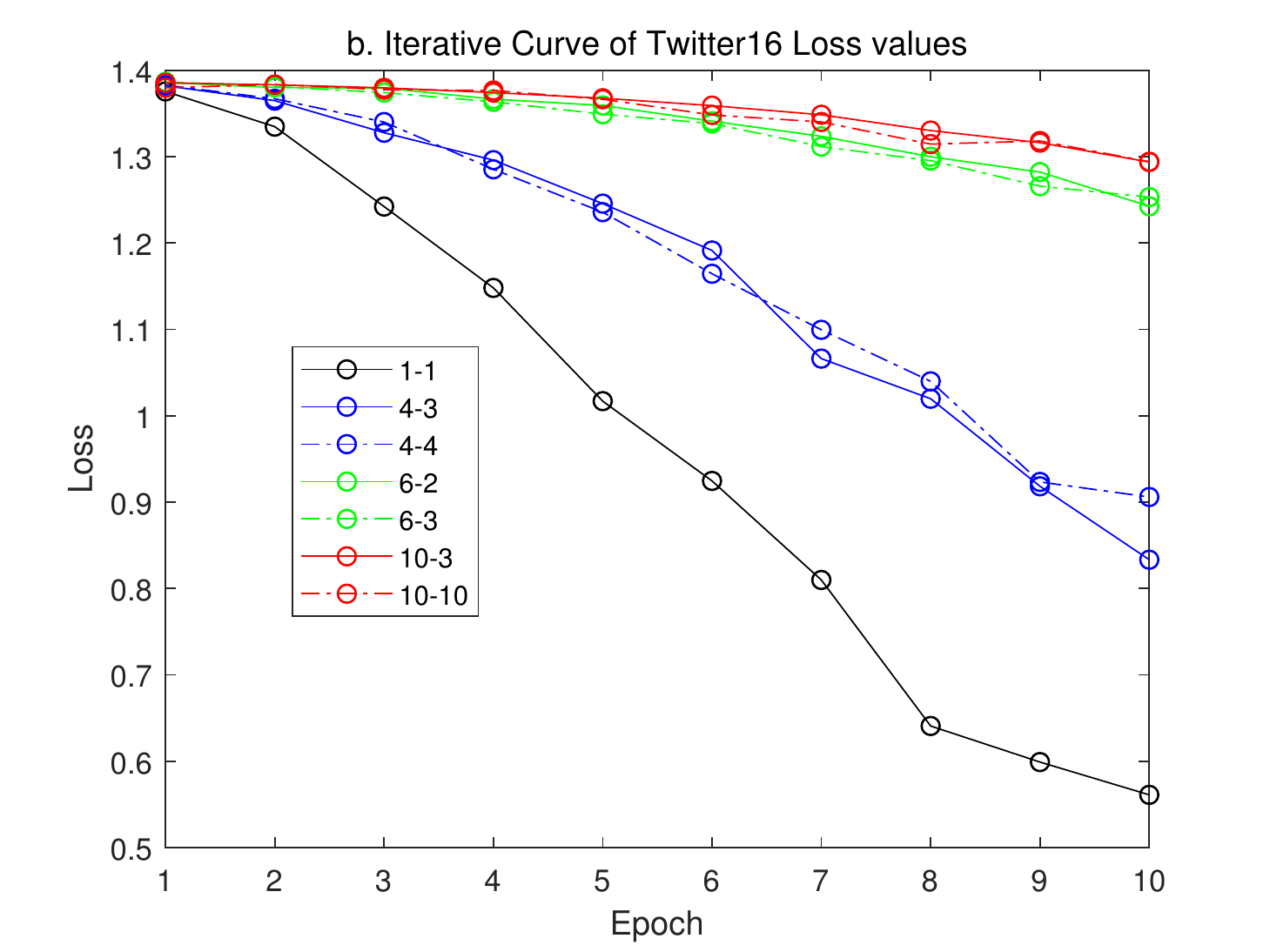}
    \end{minipage}
}
\caption{Iterative curves of loss values for Twitter15 and Twitter16 datasets.(The number in the pairs of numbers on the legend are the value of $m$ and the value of $k$ respectively)}
\label{fig:2} 
\end{figure}

Fig.\ref{fig:2}  shows the experimental results of the model with the different total number of clients  $m$ and the different number of clients $k$ chosen for each epoch, where the same color curves on these figures indicate that they have the same $m$. It can be seen from the figures that the loss values iterative curves of these two datasets have similar characteristics. The more the total number of social platforms $m$ of the client, the slower the model loss decreases. When the value of $m$ is the same, the larger the number $k$ of randomly chosen clients in each epoch, the better the performance of the model. This result is understandable, when the number of social platforms participating in the detection increases, the noise of different platforms will also increase. If the number of clients chosen for each epoch is small, it will also make model training more difficult.

The $\lambda$ in the local model is a parameter that adjusts the influence of the data features from other social platforms on its local model, so we also explore the effect of the parameter on the FedGAT model. We set $\lambda  \in (0.2,0.4,0.6,0.8,1)$ and take Twitter15 and Twitter16 as the datasets of two social platforms participating in rumor detection.

\begin{figure}[h]
\subfigure
{
    \begin{minipage}[b]{.5\linewidth}
        \includegraphics[scale=0.42]{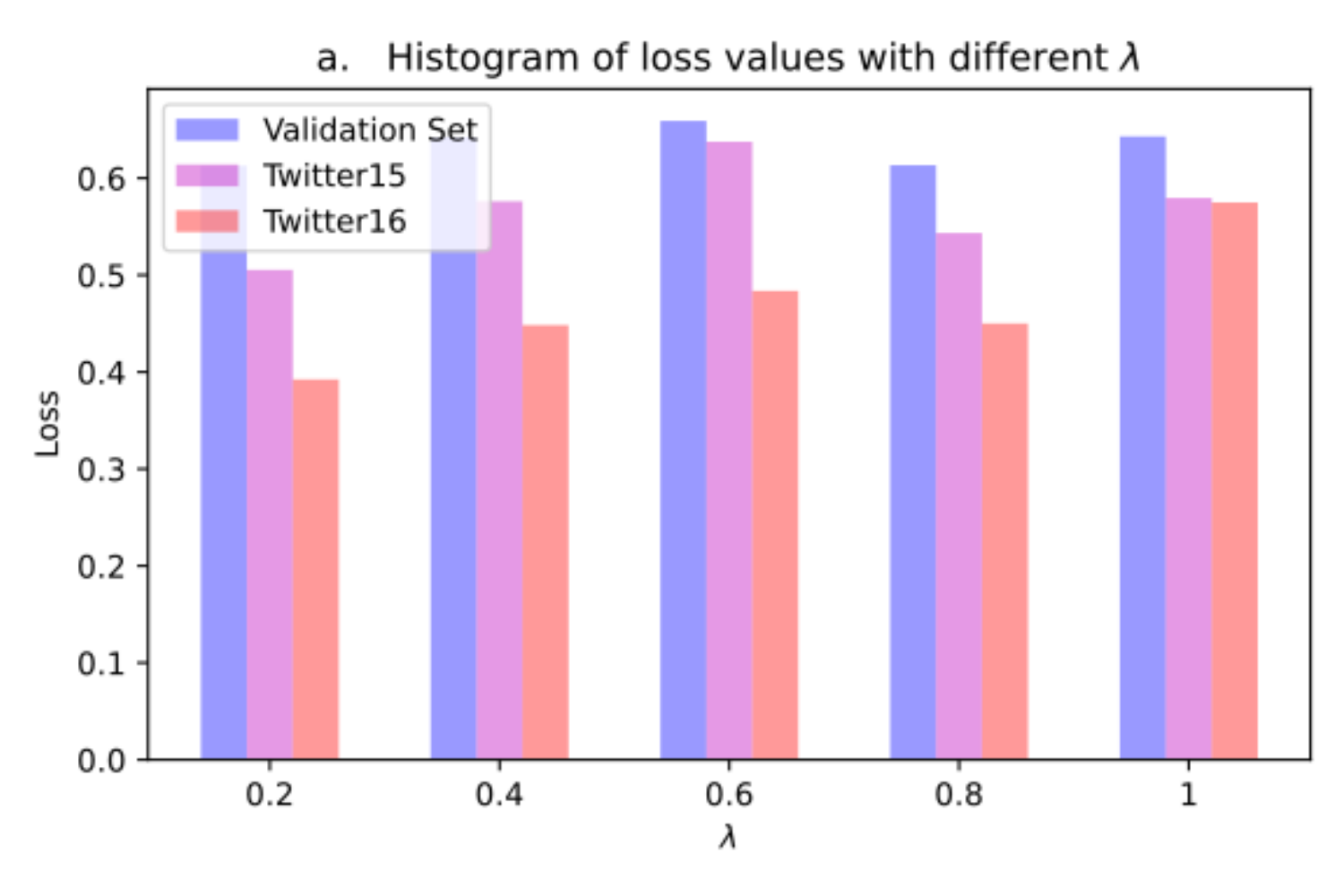}
    \end{minipage}
}
\subfigure
{
 	\begin{minipage}[b]{.5\linewidth}
        \includegraphics[scale=0.42]{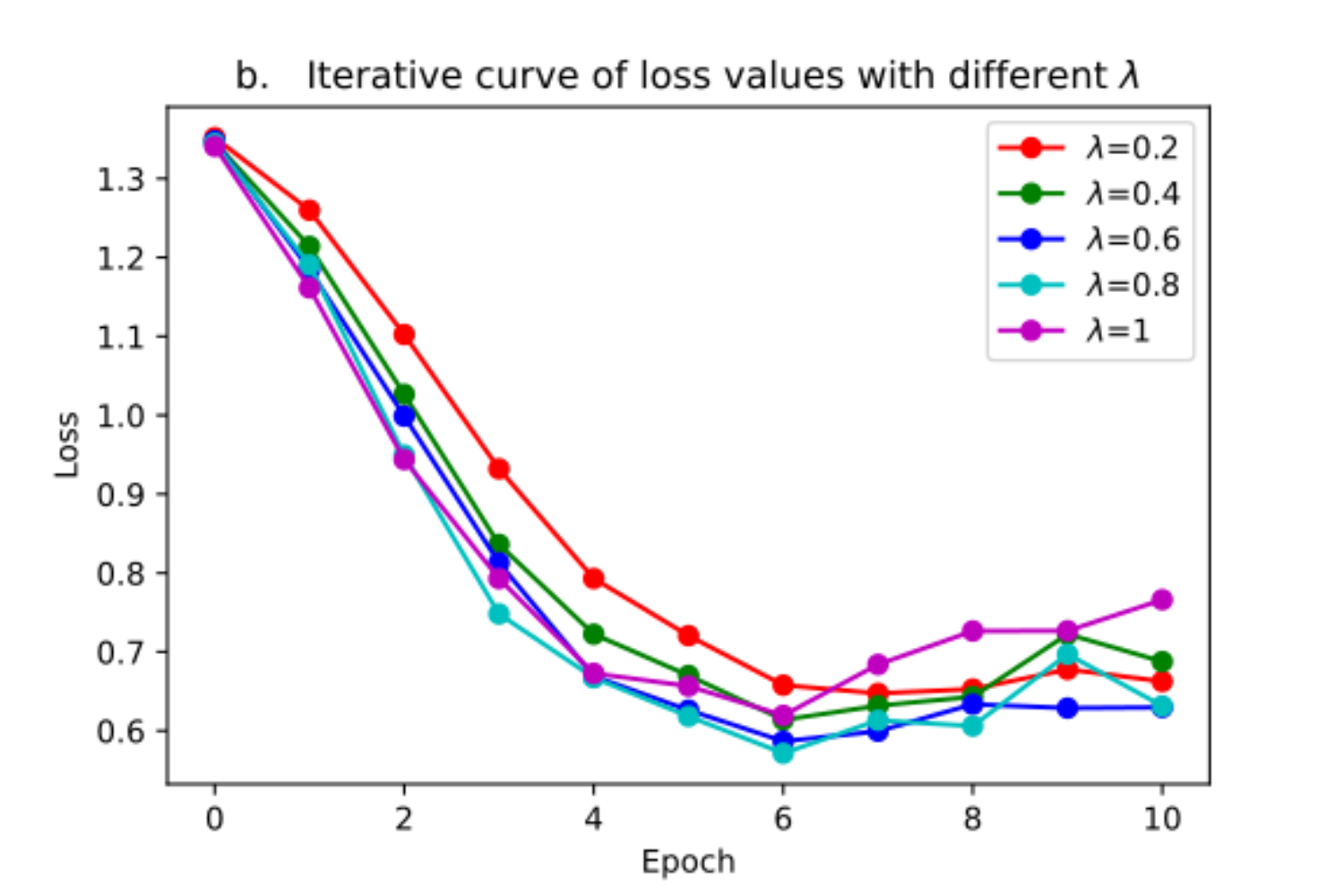}
    \end{minipage}
}
\caption{The experimental results of the loss values with different $\lambda$}
\label{fig:3} 
\end{figure}

The experiment results are shown in Fig.\ref{fig:3}, where Fig.\ref{fig:3}a shows the lowest loss value can be achieved of the server-side validation set, Twitter15 and Twitter16 datasets with different $\lambda$, and Fig.\ref{fig:3}b shows the iterative curve of loss values with different $\lambda$. From Fig.\ref{fig:3}a, it can be seen that the loss values of the three datasets with different $\lambda$ levels have no obvious regularity, and they are all in a stable loss state. From the different iterative curves in Fig.\ref{fig:3}b, it can be found that the smaller the value of $\lambda$, the more gradual the decrease of the iterative curve. So the more fully aggregated the information of other social platforms, the faster the loss value of model training can be reduced, which also proves that the establishment of a cross-platform rumor detection model is meaningful.

\section{Conclusion}\label{sec6}

The problem of rumor detection across different social platforms is an area worth researching. In this paper, we combine the federated learning framework with a bidirectional graph attention network rumor detection model to construct a federated graph attention network model. It can solve the problem of isolated data island in rumor detection on different social platforms, and can conduct cross-platform rumor detection safely and reliably. It can be seen from the simulation experiment results using the public Twitter rumor detection datasets that the FedGAT model proposed in this paper can achieve excellent results when processing datasets from different platforms at the same time, and it can be found that rumor detection is very suitable for building a model based on graph attention network. In addition, we also analyzed the influence of parameters such as $m$,$k$ and $\lambda$ in our model. We found that the minimum loss value that the model can achieve under different $\lambda$ levels is not much different, but it has an impact on the model optimization rate of the model. The larger the $\lambda$, the faster the loss value of model training can be reduced. The increase of the total number of platforms $m$ is not conducive to more accurate rumor prediction. In each rumor detection training, as much client data as possible should be used for each training epoch. 

Since the experiment in this paper only artificially constructs datasets to simulate rumor data of different social platforms, there may be some differences with the actual cross-platform rumor detection dataset. We expect the cross-platform rumor detection FedGAT model to be able to perform model optimization based on real-life multi-social platform scenarios in the future. In addition, the Bi-GAT model used in this paper can also be replaced by other existing excellent rumor detection models, and the horizontal federated learning paradigm can also be modified accordingly. We look forward to more research on rumor detection in the future.

\section*{Declarations} 

\textbf{Funding} None.\\
\textbf{Conflicts of interest/Competing interests} The authors declare that they have no conficts of interest/competing interests.\\
\textbf{Ethics approval} Not Applicable.\\
\textbf{Consent to participate} Not Applicable.\\
\textbf{Consent for publication} Not Applicable.\\
\textbf{Availability of data and material} The data is publicly available and from previous studies \cite{ma2017detect}.\\
\textbf{Code availability} The source code is available at \url{https://github.com/baichuanzheng1/FedGAT.git}.\\
\textbf{Authors' contributions} H. Wang and C. Bai contributed the central idea, analysed most of the data, and wrote the initial draft of the paper. J. Yao contributed to refining the ideas and revised the manuscript. All authors reviewed the results and approved the fnal version of the manuscript.

\end{document}